\documentclass{aa}  

\usepackage{graphicx}
\usepackage{txfonts}
\usepackage{natbib}
\usepackage[colorlinks=true,allcolors=blue]{hyperref}
\usepackage{lscape}
\usepackage{xcolor}
\usepackage{multicol}
\usepackage{multirow}
\usepackage{placeins}

\newcommand{\msun}{$M_{\odot}$}
\newcommand{\rsun}{$R_{\odot}$}
\newcommand{\mzams}{$M_{\rm ZAMS}$}
\newcommand{\snii}{SN~II}
\newcommand{\sneii}{SNe~II}
\newcommand{\eq}{\,$=$\,}
\newcommand{\ml}{\,$\pm$\,}

\newcommand{\kms}{km\,s$^{-1}$}
\newcommand{\mlow}{$M_{\rm lo}$}
\newcommand{\mhigh}{$M_{\rm up}$}

\begin{document}

   \title{Impact of stellar rotation on type II supernova progenitor masses from pre-explosion imaging}
    
   \author{L. Martinez \inst{1}
      \and O.~G. Benvenuto \inst{1,2,}\thanks{Member of the Carrera del Investigador Científico, Comisión de Investigaciones Científicas de la Provincia de Buenos Aires (CIC-PBA), Argentina.}
      \and M.~A. De Vito \inst{1,2}
           }

   \institute{
    Instituto de Astrofísica de La Plata (IALP), CCT-CONICET-UNLP. Paseo del Bosque S/N, B1900FWA, La Plata, Argentina \\ \email{laureano@fcaglp.unlp.edu.ar} 
   \and
    Facultad de Ciencias Astronómicas y Geofísicas - UNLP. Paseo del Bosque S/N, B1900FWA, La Plata, Argentina
   }

\authorrunning{L. Martinez et al.}
\titlerunning{Impact of rotation on SN~II progenitor masses}

\date{Received XXX; accepted XXX}


\abstract 
{
The initial masses of red supergiant (RSG) type II supernova (\snii) progenitors are commonly inferred from pre-explosion imaging by converting the progenitor luminosity into an initial mass estimate using non-rotating stellar evolution models.
However, stellar rotation affects the evolution and may influence these estimates.
We investigate how the observed distribution of rotational velocities in massive stars influences the progenitor initial masses of \sneii\ inferred from pre-SN imaging.
We compare initial mass estimates obtained from non-rotating models with those derived from rotating models, where the initial rotational velocities of the stellar models are sampled from the observed distribution.
We analyse the inferred progenitor initial masses by (i) comparing the results for each SN individually, (ii) examining the overall probability density function, (iii) constructing the cumulative distribution function, and (iv) determining the upper initial-mass boundary.
In all cases, the distributions obtained from rotating models are slightly shifted towards lower masses, although the differences remain smaller than the typical uncertainties.
When using the observed distribution of initial rotational velocities for massive stars, we infer an upper initial-mass limit for \snii\ progenitors of 20.4$^{+2.3}_{-1.9}$\,\msun.
Taken together, these analyses demonstrate that stellar rotation has only a modest impact on progenitor mass estimates from pre-SN imaging within the current observational and model uncertainties when the observed distribution of initial rotational velocities is taken into account. Therefore, adopting this distribution leads to small differences compared to non-rotating models.
}


\maketitle
\nolinenumbers

\section{Introduction}
\label{sec:intro}

Type~II supernovae (\sneii\footnote{We use `SNe II' to refer to hydrogen-rich core-collapse SNe, excluding type IIb, IIn, and SN~1987A-like events.}) are the explosive deaths of massive stars ($\gtrsim$\,9\,\msun) through the core-collapse mechanism, characterized by prominent hydrogen features in their spectra \citep{minkowski41,filippenko+93}.
The presence of such hydrogen features indicates that the progenitor stars retained a substantial fraction of their hydrogen envelopes throughout their evolution up to the core-collapse event.
This scenario has been further supported by the identification of red supergiant (RSG) stars in archival pre-explosion images at the locations of \snii\ events, providing direct evidence that RSGs constitute the progenitors of \sneii\ \citep{smartt15,vandyk25}.

\sneii\ are the most common type of core-collapse SN (CCSN) in nature \citep{shivvers+17,pessi_t+25}.
However, the exact ranges of the physical properties of their progenitors remain uncertain, particularly their initial masses.
Several methods have been used to estimate progenitor initial masses, including nebular spectral modelling \citep[e.g.][]{jerkstrand+15, fang+25} and light-curve modelling \citep[e.g.][]{morozova+18, martinez+20}, both of which link post-explosion observables to progenitor mass through inference methods that rely on stellar evolution models.
Additionally, the study of stellar populations at SN explosion sites \citep{maund17, williams+18} also provide constraints on progenitor masses.

While the previous approaches rely on post-explosion observables or stellar population studies, a more direct method to estimate progenitor properties is to detect individual progenitors in pre-explosion images.
This allows the measurement of their luminosity and effective temperature prior to the SN event.
The progenitor luminosity is obtained from fits to the spectral energy distribution or using bolometric corrections \citep[e.g.][]{davies+18, vandyk+19}.
The next step is to convert the pre-explosion progenitor luminosity into an initial mass estimate using stellar evolution models. 

The estimation of the initial masses of observed \snii\ progenitors is generally consistent with estimates obtained from post-explosion techniques, such as light-curve modelling and the analysis of nebular spectra \citep[e.g.][]{jerkstrand+15, eldridge+19a}.
In particular, \citet{eldridge+19a} produced SN light-curve models using progenitors computed with the STARS stellar evolution code \citep{eldridge+04}, the same code that has been used to link observed progenitor luminosities to initial masses \citep{smartt15}, finding consistent initial-mass estimates from the two approaches.

Estimating the initial mass from the pre-explosion progenitor luminosity has typically relied on publicly available grids of stellar evolution models.
These are typically computed either without rotation or assuming a fixed fraction of the critical velocity at the zero-age main sequence (ZAMS), most commonly 
\(\Omega_{\rm ini}/\Omega_{\rm crit}=0.4\) (where $\Omega_{\rm ini}$ and $\Omega_{\rm crit}$ are the initial and critical surface angular velocities, respectively), using codes such as \texttt{KEPLER} \citep{woosley+07}, \texttt{STARS} \citep{eldridge+08}, \texttt{GENEVA} \citep{ekstrom+12}, and \texttt{MESA} \citep{paxton+11} through the \texttt{MIST} project \citep{choi+16}. 
Hence, the available rotating grids usually represent only a single fixed value rather than a range of initial rotational velocities.
However, stellar rotational velocities at the ZAMS span a wide range \citep{huang+10, holgado+22}.

Rotation has a major impact on the evolution of massive stars.
While rotation enhances wind mass-loss rates through centrifugal effects \citep{maeder+00, langer12}, its primary effect is the enhancement of internal chemical mixing. Rotational instabilities transport fresh fuel into the core, thereby extending the duration of the nuclear burning phases and increasing core masses, while also affecting surface abundances and nucleosynthetic yields, among other effects \citep{maeder+00}.
Variations in the rotation rate alter the amount of internal mixing and therefore the mass of the helium core. Because the helium core mass sets the final stellar luminosity, rotation ---via enhanced mixing--- changes the initial mass to final luminosity relation.

In this work, we estimate the initial masses of \snii\ progenitors from their pre-explosion luminosities by explicitly incorporating the observed distribution of stellar rotational velocities at the ZAMS.
Our main goal is to quantify the differences between this approach and the traditional method based on non-rotating models, in order to assess the impact of the observed initial velocity distribution on the inferred population of \snii\ progenitors.

The current paper is organised as follows.
In Sect.~\ref{sec:sample}, we provide a brief description of the \snii\ sample. 
Section~\ref{sec:methods} presents the grids of stellar evolution models and the methodology used to estimate the progenitor initial-mass distributions. 
The results obtained with non-rotating models and with rotating models based on the observed initial rotational velocities of massive stars are presented in Sect.~\ref{sec:results}. 
In Sect.~\ref{sec:analysis}, we perform statistical comparisons between the initial-mass distributions obtained from results using rotating and non-rotating models. 
In Sect.~\ref{sec:discussion}, we discuss key implications of our analysis, such as the nature of low-luminosity progenitors, the current statistical significance of the RSG problem, and the potential role of binary interaction channels.
Finally, Sect.~\ref{sec:conclusions} summarises our main conclusions. 
Additional figures not included in the main body of the manuscript are provided in the Appendix.

\section{Sample of \snii\ progenitors}
\label{sec:sample}

\begin{table}
\caption{Pre-SN luminosities of \snii\ progenitors in the sample analysed in this work.}
\label{table:logL}
\centering
\begin{tabular}{lcc}
\hline\hline\noalign{\smallskip}
SN & $\log{L/L_{\odot}}$ & Reference \\
\hline\noalign{\smallskip}
2004A    & 4.90 $\pm$ 0.10  &  \citet{davies+18} \\
2004et   & 4.77 $\pm$ 0.07  &  \citet{davies+18} \\
2006my   & 4.97 $\pm$ 0.18  &  \citet{davies+18} \\
2008bk   & 4.53 $\pm$ 0.07  &  \citet{davies+18} \\
2008cn   & 5.10 $\pm$ 0.07  &  \citet{davies+18} \\
2009hd   & 5.24 $\pm$ 0.08  &  \citet{davies+18} \\
2009ib   & 5.12 $\pm$ 0.14  &  \citet{takats+15} \\
2009kr   & 5.13 $\pm$ 0.23  &  \citet{davies+18} \\
2009md   & 4.50 $\pm$ 0.20  &  \citet{davies+18} \\
2012A    & 4.57 $\pm$ 0.09  &  \citet{davies+18} \\
2012aw   & 4.92 $\pm$ 0.12  &  \citet{davies+18} \\
2012ec   & 5.16 $\pm$ 0.07  &  \citet{davies+18} \\
2013ej   & 4.69 $\pm$ 0.07  &  \citet{davies+18} \\
2017eaw  & 5.08 $\pm$ 0.07  &  \citet{vandyk+19} \\
2018zd   & 4.50 $-$   5.10  &  \citet{Hiramatsu+21b} \\
2018aoq  & 4.72 $\pm$ 0.12  &  \citet{oneill+19} \\
2022acko & 4.30 $-$   4.50  &  \citet{vandyk+23b} \\
2023ixf  & 4.95 $\pm$ 0.07  &  \citet{vandyk+24} \\
2024ggi  & 4.92 $\pm$ 0.05  &  \citet{xiang+24b} \\
2024abfl & 4.32 $-$   4.80  &  \citet{luo+25} \\
2025pht  & 5.00 $\pm$ 0.09  &  \citet{kilpatrick+25} \\
\hline
\end{tabular}
\end{table}

In this work, we aim to determine the differences in \snii\ progenitor initial masses inferred from pre-SN imaging when constraining with rotating and non-rotating models.
In this context, the key observable is the pre-SN luminosity, which is converted into an initial mass through stellar evolution models.

We therefore collected a sample of \snii\ progenitors detected in pre-SN images with direct luminosity estimates, excluding those with only upper limits.
Based on this criterion, we initially gathered 24~\sneii\ from the literature, of which some were discarded.
Two of the progenitors we excluded are those of SN~2003gd and SN~2005cs. These stars have luminosities estimated only from the \textit{Hubble} Space Telescope F814W band, which may significantly underestimate their true values \citep{beasor+25}.
In addition, the progenitor of SN~2020jfo remains uncertain \citep{vandyk25}, as the fit to its observed spectral energy distribution is consistent with a star below the mass range expected to explode as either a CCSN or an electron-capture SN (ECSN), even when assuming a dust-obscured RSG model \citep{kilpatrick+23b}.
Therefore, our final sample consists of 21~\snii\ progenitors, listed in Table~\ref{table:logL} together with their pre-SN luminosities.

\section{Methods}
\label{sec:methods}

In this section, we describe the methods used to estimate the initial masses of \snii\ progenitors.
Section~\ref{subsec:models} presents the computation of stellar evolution models for rotating massive stars and the impact of rotation on their pre-SN luminosities.
Section~\ref{subsec:velo_distribution} details the observationally derived distribution of initial rotational velocities for massive stars.
Finally, Sect.~\ref{subsec:mc} describes the Monte Carlo (MC) approach used to combine the observed progenitor luminosities and rotational velocities with the model grids to infer the progenitor initial-mass distributions.

\subsection{Modelling rotating massive stars}
\label{subsec:models}

\begin{table}
\caption{Maximum initial mass across the computed grids for different initial surface velocities.}
\label{table:max_initial_mass}
\centering          
\begin{tabular}{cc}
\hline\hline\noalign{\smallskip}
$v_{\rm ini}$ & Maximum \mzams\ \\
(\kms) & (\msun) \\
\hline\noalign{\smallskip}
 0 $-$ 300 & 30 \\
 350 & 27 \\
 400 & 23 \\
 450 & 20 \\
 500 $-$ 550 & 17 \\
 600 & 16 \\
\hline
\end{tabular}
\end{table}

\begin{table}
\caption{Properties of non-rotating stellar models at the end of core carbon burning.}
\label{table:models_vsurf_0}
\centering          
\begin{tabular}{cccccccc}
\hline\hline\noalign{\smallskip}
\mzams\ & $\log(L/L_{\odot})$ & $R$ & $M_{\rm final}$ & $M_{\rm env}$ & $M_{\rm He}$ & $M_{\rm CO}$ \\
(\msun) & & (\rsun) & (\msun) & (\msun) & (\msun) & (\msun) \\
\hline\noalign{\smallskip}
 9 & 4.52 & 418 &  8.94 &  6.47 &  2.47 &  1.37 \\
10 & 4.63 & 480 &  9.95 &  7.11 &  2.84 &  1.59 \\
11 & 4.72 & 536 & 10.95 &  7.75 &  3.21 &  1.80 \\
12 & 4.80 & 586 & 11.95 &  8.44 &  3.51 &  2.02 \\
13 & 4.87 & 640 & 12.92 &  9.02 &  3.90 &  2.29 \\
14 & 4.94 & 694 & 13.88 &  9.57 &  4.31 &  2.59 \\
15 & 5.00 & 746 & 14.84 & 10.14 &  4.69 &  2.89 \\
16 & 5.07 & 804 & 15.85 & 10.66 &  5.20 &  3.27 \\
17 & 5.12 & 850 & 16.80 & 11.24 &  5.56 &  3.58 \\
18 & 5.15 & 879 & 17.64 & 11.82 &  5.82 &  3.83 \\
19 & 5.20 & 932 & 18.51 & 12.25 &  6.26 &  4.23 \\
20 & 5.25 & 1000 & 19.26 & 12.45 &  6.81 &  4.71 \\
21 & 5.31 & 1070 & 19.93 & 12.55 &  7.38 &  5.22 \\
22 & 5.35 & 1124 & 20.61 & 12.72 &  7.88 &  5.68 \\
23 & 5.37 & 1149 & 21.22 & 12.84 &  8.38 &  6.14 \\
24 & 5.42 & 1227 & 21.63 & 12.66 &  8.97 &  6.68 \\
25 & 5.46 & 1294 & 21.96 & 12.43 &  9.53 &  7.20 \\
26 & 5.49 & 1322 & 22.34 & 12.31 & 10.03 &  7.66 \\
27 & 5.52 & 1387 & 22.52 & 11.94 & 10.58 &  8.16 \\
28 & 5.55 & 1441 & 22.78 & 11.70 & 11.08 &  8.63 \\
29 & 5.59 & 1508 & 22.72 & 11.07 & 11.64 &  9.15 \\
30 & 5.62 & 1564 & 22.68 & 10.50 & 12.18 &  9.65 \\
\hline
\end{tabular}
\tablefoot{Columns: (1) Initial mass; (2) logarithm of the effective temperature; (3) logarithm of the stellar luminosity; (4) stellar radius; (5) total final mass; (6) hydrogen-rich envelope mass; (7) helium-core mass; (8) carbon-oxygen core mass.}
\end{table}

\begin{figure*}
\centering
\includegraphics[width=0.495\textwidth]{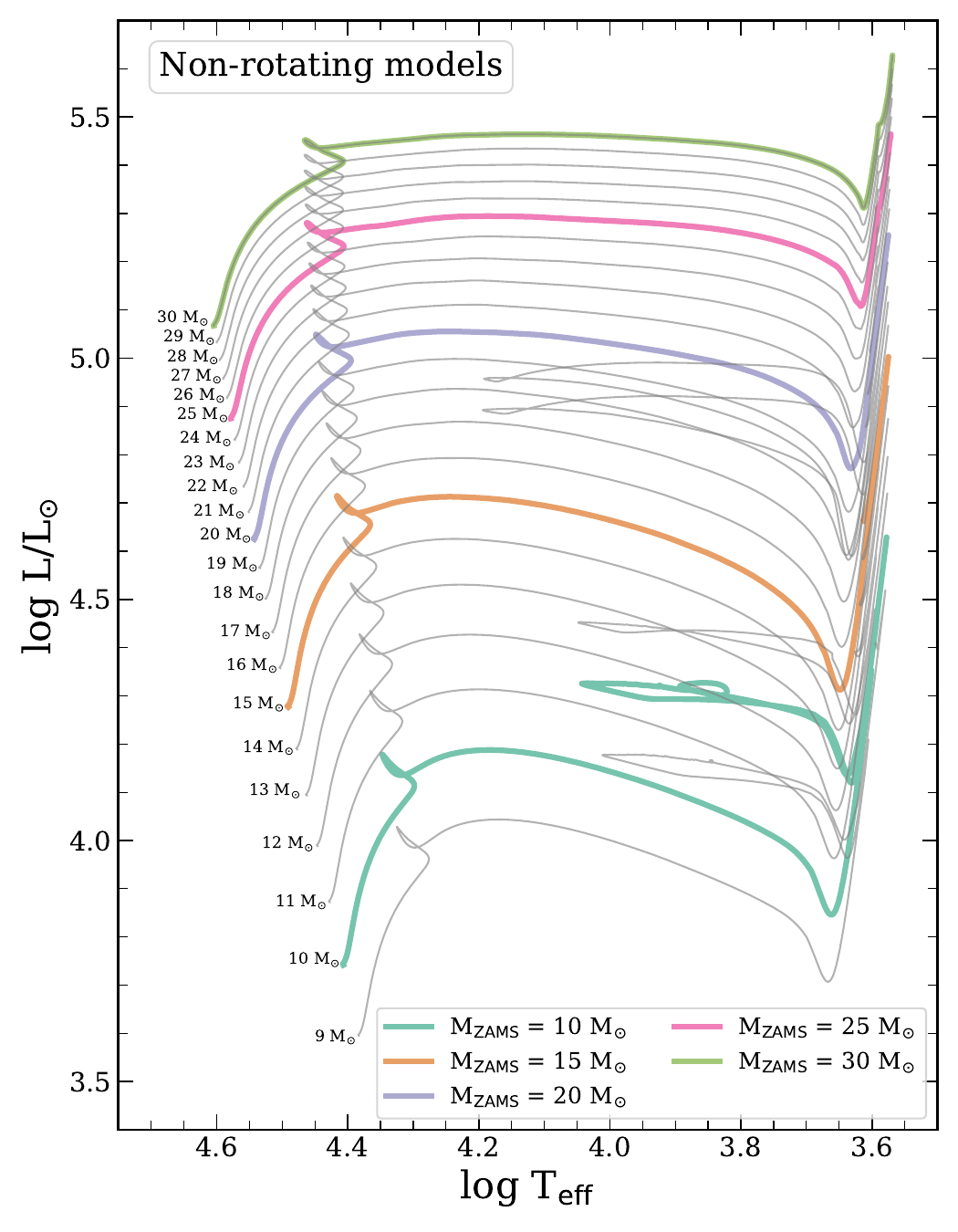}
\includegraphics[width=0.495\textwidth]{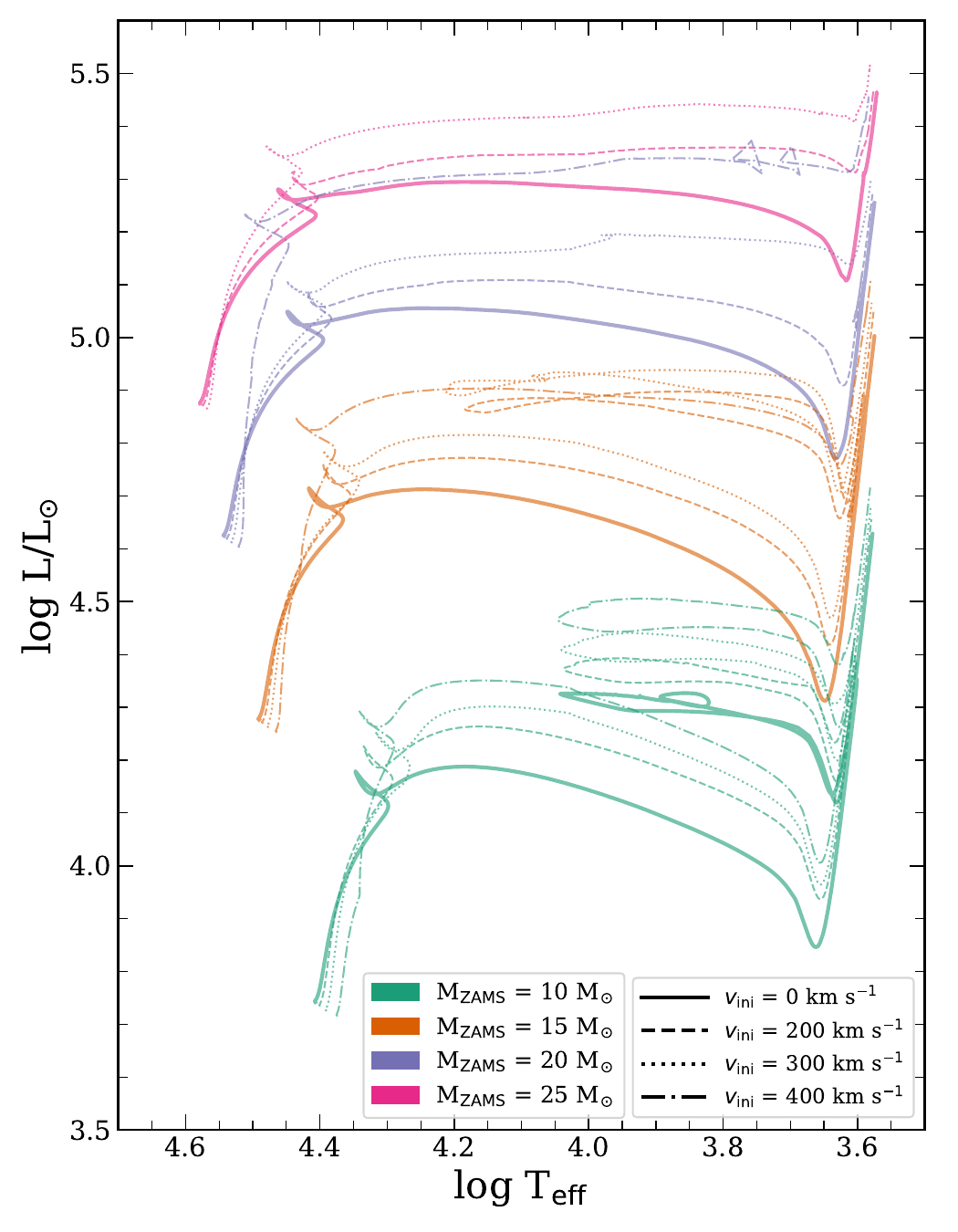}
\caption{HR diagrams for a subsample of stellar models. The left panel shows only non-rotating models. Coloured lines correspond to models with initial masses of 10, 15, 20, and 25\,\msun, while the remaining models are shown as faint grey lines. The right panel presents evolutionary tracks for stars with initial masses of 10, 15, 20, and 25\,\msun\ and different initial rotational velocities as indicated in the panel label. We note that the model with an initial mass of 25\,\msun\ and an initial rotational velocity of 400\,\kms\ lies outside the parameter space of the current study (see Table~\ref{table:max_initial_mass}).}
\label{fig:hr}
\end{figure*}

\begin{figure}
\centering
\includegraphics[width=0.5\textwidth]{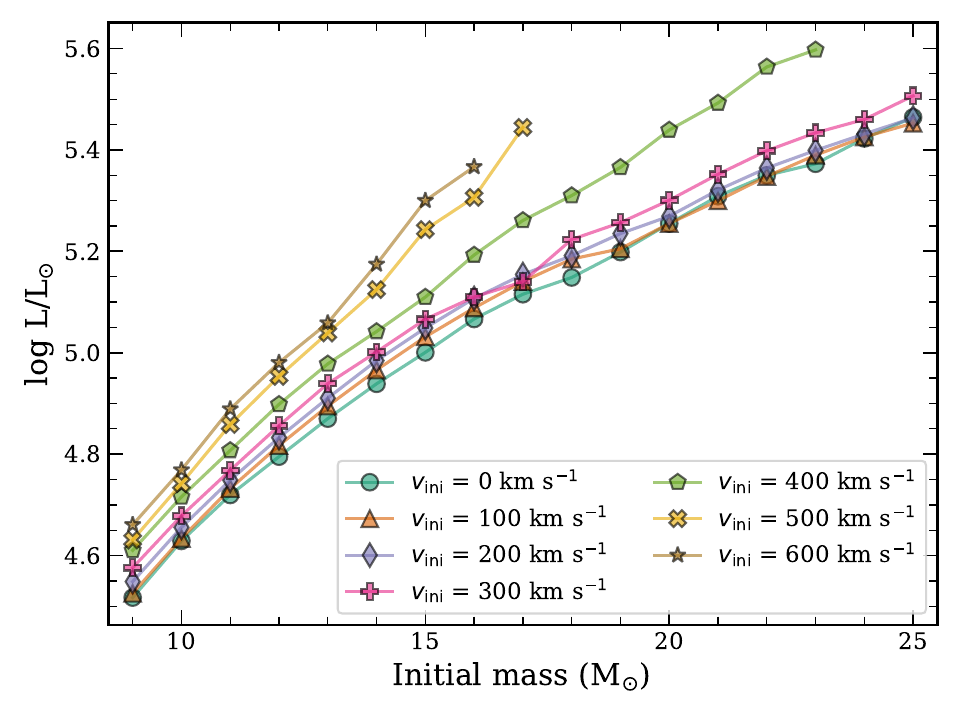}
\caption{Stellar luminosities at core carbon depletion for a range of initial masses and surface velocities. Only a subset of models is shown for visualisation purposes.}
\label{fig:logL_models}
\end{figure}

To estimate progenitor initial masses, it is necessary to compare the observed pre-SN luminosity with that predicted by stellar evolution models.  
For this purpose, we computed evolutionary models of single massive stars at solar metallicity \citep[Z$_{\odot}$\eq0.0154;][]{asplund+21} with the public stellar evolution code \texttt{MESA}\footnote{\url{https://docs.mesastar.org/en/latest/}}, version 24.08.1 \citep{paxton+11,paxton+13,paxton+15,paxton+18,paxton+19,jermyn+23}.
Each stellar model was evolved from the ZAMS until core carbon depletion, which we defined as the stage when central carbon mass fraction drops below 0.01.

Wind mass loss rates for massive stars were computed following the latest prescriptions: \citet{bjorklund+23} for hot stars, and \citet{wen+24} for cool stars.
In the transition region, the wind mass loss rate is a combination of both prescription using the same method as in the `Dutch' wind scheme defined in \texttt{MESA}.
We adopt the wind prescriptions from \citet{wen+24}, which were derived for RSGs in the Large Magellanic Cloud, even though our stellar models assume solar metallicity. This is justified because previous studies indicate that RSG wind mass-loss rates show little dependence on metallicity, with no strong correlation between the two \citep{goldman+17, antoniadis+25}.

We used the Ledoux criterion to determine the convective regions and set the mixing length parameter to $\alpha_{\rm mlt}\eq2.0$.
The semiconvection and thermohaline mixing efficiency parameters were both set to 1.0 \citep{kippenhahn+80, langer+83}. 
Convective-core overshooting was treated with the step formalism during hydrogen- and helium-core burning, adopting overshooting parameters of $\alpha_{\rm os}$\eq0.15 \citep{martins+13} and 0.03 \citep{li+19} pressure scale heights, respectively. 
The hydrogen-core overshooting parameter was calibrated by \citet{martins+13} through comparisons with the observed width of the main sequence using rotating stellar models with different overshooting prescriptions. While this calibration partially accounts for the additional mixing induced by rotation, it is based on models computed with a single initial rotational velocity (or a fixed fraction of the critical velocity), rather than a distribution of rotation rates.
During carbon burning, we adopted the exponential approach implemented in \texttt{MESA} to account for convective overshooting, with a parameter $f$\eq0.004 \citep{farmer+16}.

There are currently two approaches to treat chemical mixing and angular momentum transport caused by rotationally induced instabilities in one-dimensional stellar evolution models: the diffusion approximation \citep{endal+78, heger+00} and the diffusion--advection approach \citep{zahn92}.  
\texttt{MESA} adopts the former \citep{paxton+13}.
The rotationally induced instabilities included in our models are: dynamical shear instability, Solberg--H{\o}iland instability, secular shear instability, Eddington$-$Sweet circulation, and Goldreich--Schubert--Fricke instability \citep[see][for details]{heger+00}. Additionally, we included the effects of internal magnetic fields following the Spruit--Tayler dynamo approach \citep{spruit02, heger+05, paxton+13}.
In the diffusion approximation, two free parameters require calibration: $f_{c}$, the ratio of the turbulent viscosity to the diffusion coefficient, and $f_{\mu}$, which describes the sensitivity of rotationally induced mixing to $\mu$-gradients. 
We adopted $f_{c}$\eq1/30 and $f_{\mu}$\eq0.05, following \citet{heger+00}.

Using the stellar evolution parameters described above, we computed models varying the initial mass and the initial rotational surface velocity.  
The initial surface velocities were chosen in the range 0–600\,\kms, with steps of 50\,\kms. The upper limit is motivated by the observed velocity distribution of massive stars (see Sect.~\ref{subsec:velo_distribution}).  
The models were assumed to rotate uniformly on the ZAMS.

The minimum initial mass in our grids is set by the lowest-mass star capable of producing a CCSN at the end of its evolution.
We adopt neon ignition ---either central or off-centre--- as the criterion defining this limit, since stars that ignite neon are generally expected to undergo core collapse rather than end as oxygen--neon (ONe) white dwarfs or ECSNe \citep{nomoto+84, kippenhahn+13_book, limongi+24}.
We note that ECSNe are expected to arise from slightly lower initial masses and would observationally appear as \sneii, which are the events considered in this work. However, ECSNe are predicted to occur over a relatively narrow initial-mass range, of about 0.5\,\msun\ \citep{poelarends+08, limongi+24}. Including this channel could slightly shift the inferred initial-mass distribution towards lower masses, but it is not expected to significantly affect our results.
Our models place the minimum initial mass for producing a \snii\ explosion at 9\,\msun\footnote{We note that our simulations are stopped at the end of core carbon burning. For the lowest-mass models, we continued the evolutionary calculations in order to determine whether neon ignition occurs and thus assess the final fate of the star.}. 
Therefore, we set the minimum initial mass to 9\,\msun, independent of the initial surface rotational velocity.

The maximum initial mass depends on the initial surface velocity. For grids with velocities between 0 and 300\,\kms, it is 30\,\msun.  
Within this velocity range, initially more massive stars still retain a significant hydrogen-rich envelope at the end of evolution and are therefore expected to explode as \sneii. However, their final luminosities exceed those inferred for observed \snii\ progenitors, and we do not consider them further.  
For higher initial surface velocities, the maximum initial mass is set by the condition that the star retains a significant hydrogen-rich envelope at the end of its evolution, while more massive stars lose almost their entire envelope and would not end their lives as \sneii.
This dependency arises because rotation affects stellar evolution through enhanced mass loss and chemical mixing \citep{maeder09}. 
Increased mixing supplies more hydrogen to the core, producing more massive helium cores.  
Consequently, these stars are more luminous and develop stronger winds, which act together to the intrinsic enhanced mass loss to reduce the hydrogen-rich envelope mass.   
As a result, the maximum initial mass decreases with increasing initial surface velocity.
The maximum initial-mass values for each initial velocity are listed in Table~\ref{table:max_initial_mass}.

Figure~\ref{fig:hr} shows Hertzsprung--Russell (HR) diagrams for a subsample of our models.
The left panel presents evolutionary tracks for non-rotating models, while the right panel shows tracks for rotating 10, 15, 20, and 25\,\msun\ models at different initial rotational velocities, illustrating the effects of stellar rotation.
Additionally, Figs.~\ref{fig:hr_all1} and \ref{fig:hr_all2} show HR diagrams of models spanning the entire range of initial masses and a selected set of initial rotational velocities.

The evolution of rotating models follows the expected behaviour relative to non-rotating models.
On the ZAMS and during the early stages of the main-sequence evolution, rotation induces a shift of the tracks towards lower luminosities and effective temperatures, primarily due to the reduction of the effective gravity.
As evolution proceeds, the evolutionary tracks of rotating models become more luminous than those of non-rotating models.
This results from rotational mixing during the main-sequence evolution. On the one hand, enhanced mixing supplies fresh hydrogen to the hydrogen-burning convective core, thereby producing more massive helium cores and more luminous stars.
On the other hand, rotational mixing transports helium and other hydrogen-burning products to the surface, reducing the opacity and contributing to an increase in stellar luminosity.
Owing to the more massive helium core, the post-main sequence evolution resembles that of an initially more massive star, resulting in higher luminosities compared to non-rotating models. 
For further details on the evolution of rotating stars the reader is referred to, for example \citet{meynet+00} and \citet{chieffi+13}.

In the current work, we focus primarily on the final luminosity of each model, as it will be compared with the observed progenitor luminosity to estimate the initial mass.  
Figure~\ref{fig:logL_models} shows the final luminosity for a subset of models. As expected, more massive stars reach higher luminosities at the end of their evolution.  
Additionally, due to the rotational effects described above, more rapidly rotating stars attain higher final luminosities at a given initial mass.
We note that, at a fixed final luminosity, the range of initial masses increases toward higher luminosities. For instance, at $\log L = 4.8$, the inferred initial mass spans approximately 2\,\msun, across the considered rotation rates. This spread increases to about 3\,\msun\ at $\log L = 5.0$, and to roughly 8\,\msun\ at $\log L = 5.5$, implying larger uncertainties at higher luminosities.
At low and intermediate luminosities, the range of inferred initial masses is comparable to that found by \citet{eldridge+04}, who showed that similar mass spreads arise when considering models with and without convective core overshooting. 
This highlights that uncertainties in stellar evolution modelling, such as chemical mixing processes, play an important role in setting the spread of inferred progenitor initial masses. Additionally, \citet{eldridge+04} found comparable mass spreads resulting from variations in the initial metallicity.
We note that our analysis is restricted to solar-metallicity models.
Some stellar properties at core carbon depletion for the non-rotating models are presented in Table~\ref{table:models_vsurf_0}.

\subsection{Initial rotational velocity distribution of massive stars}
\label{subsec:velo_distribution}

\begin{figure}
\centering
\includegraphics[width=0.49\textwidth]{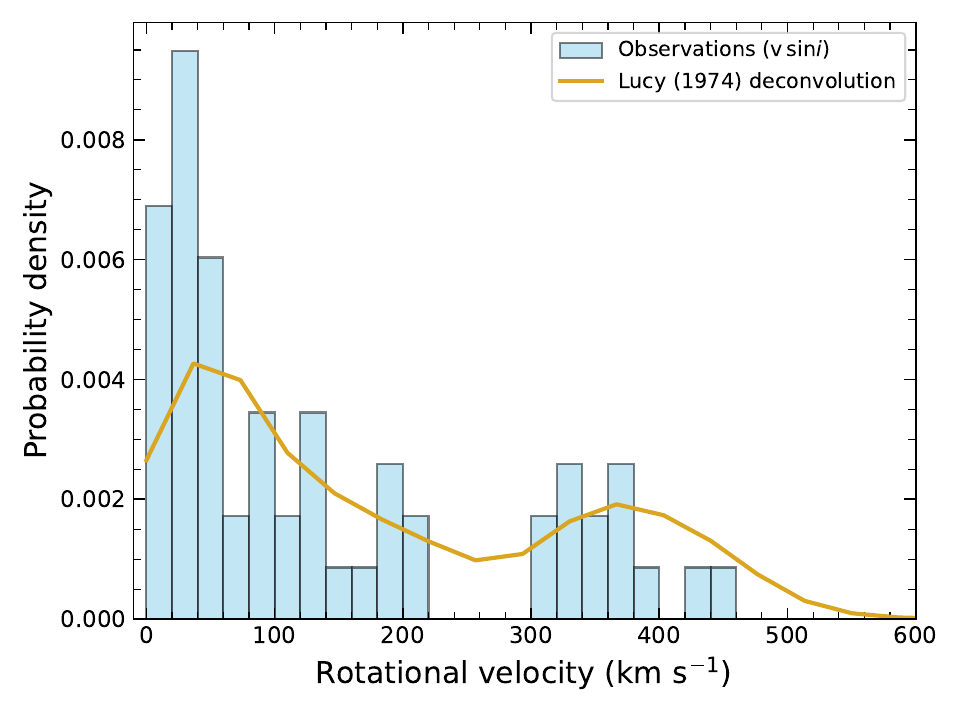}
\caption{Probability density of rotational velocities. Projected rotational velocities of young Galactic O-type stars from \citet{holgado+22} with masses below 32\,\msun\ are shown as a histogram. The deconvolved distribution, obtained using the method of \citet{lucy74}, is overplotted as a thick line.}
\label{fig:velo_distribution}
\end{figure}

When using the initial mass–final luminosity relation to infer the mass of a SN progenitor, previous studies typically rely on either non-rotating models or rotating models computed with an initial surface velocity set to a fixed fraction of the critical value \citep[e.g.][]{davies+18, davies+20a}.
In the latter case, a value of \(\Omega_{\rm ini}/\Omega_{\rm crit}=0.4\) is commonly adopted, motivated by the peak in the observed rotational velocity distribution of young B-type stars \citep{huang+10}.  
In this work, rather than assuming that all stars initially rotate at a fixed fraction of the critical velocity, we account for the full observed distribution of initial rotational velocities for massive stars at the ZAMS.

For this purpose, we used the projected rotational velocities ($v\sin i$) of single Galactic O-type stars presented by \citet{holgado+22}.  
In particular, we adopted the subsample defined in their study corresponding to stars with initial masses below 32\,\msun, which is consistent with the mass range of our model grid.
In addition, we restricted the sample to stars with surface gravities $\log g$\,$>$\,3.7\,dex in order to focus on young stars (see Fig.~8c of \citealt{holgado+22}), thereby approximating a rotational velocity distribution at the ZAMS.
Given that \citet{holgado+22} reported projected rotational velocities, we applied the deconvolution algorithm of \citet{lucy74} to recover the intrinsic rotational velocity distribution.  
Figure~\ref{fig:velo_distribution} compares the observed projected distribution with the deconvolved distribution.
The latter extends up to $\sim$600\,\kms, which justifies the choice of the maximum surface velocity adopted in our model grid (Section~\ref{subsec:models}).

\subsection{Mass distribution inference}
\label{subsec:mc}

In this section, we describe the method used to infer the initial-mass distribution of each \snii\ progenitor in the sample, combining the observed progenitor luminosity with the initial rotational surface velocity.
For each \snii\ progenitor in our sample, we draw a luminosity value from a Gaussian distribution centred on the observed luminosity, with a standard deviation equal to the reported uncertainty (see Table~\ref{table:logL}). 
In cases where the luminosity is provided as a range rather than as a value with an associated uncertainty, we adopt the midpoint of the range as the mean of the Gaussian, and set the distance between the midpoint and either boundary of the range equal to three standard deviations. 
We then sample a value from the initial rotational velocity distribution described in Sect.~\ref{subsec:velo_distribution}.

In principle, the effective temperatures of observed \snii\ progenitors could also be compared with those predicted by stellar models. However, we do not include the effective temperature as an additional parameter to infer the progenitor initial mass for the reasons discussed below.
On the one hand, the vast majority of our stellar models end their evolution with very similar effective temperatures; therefore, the effective temperature does not provide an additional constraint for the initial-mass estimate.
Only a few models deviate significantly from this value, corresponding to stars that evolve blueward due to their reduced hydrogen envelopes. At the same time, these models exhibit final luminosities that exceed those inferred for observed \snii\ progenitors, and therefore do not contribute to the inferred initial-mass distribution.
On the other hand, the uncertainties in the observed effective temperatures are typically large enough to cover the narrow interval of final effective temperatures predicted by the models.
For these reasons, we do not use the effective temperature as an additional constraint in our analysis.

With the sampled progenitor luminosity and initial surface velocity, we used our model grids to infer the corresponding initial mass. 
We first identified the closest models in the grid to the input values using the \texttt{cKDTree} algorithm from the \texttt{SciPy} \texttt{Python} package, which implements a $k$-dimensional tree to efficiently organise points in a multidimensional parameter space. 
For the nearest models, we assigned weights based on a Gaussian kernel centred on the target parameters, with the distance in parameter space determining the weight.
Finally, the progenitor’s initial mass was obtained by interpolating the models according to these weights. 
If the input progenitor luminosity is below the minimum in our models, we set the progenitor initial mass to the minimum value in the grid to avoid extrapolating beyond the model range. 

In summary, this procedure relates the sampled luminosity and initial surface velocity values to an initial mass estimate through multidimensional interpolation of the  rotating stellar models. 
We hereafter refer to this approach as the distribution-based rotating (DBR) models.
We repeated this MC procedure 10\,000 times for each \snii\ progenitor, thereby determining the posterior probability distribution of its initial mass.

\section{Results}
\label{sec:results}

\begin{table}
\caption{Progenitor initial masses from non-rotating and DBR models for each \snii\ in the sample.}
\label{table:results}
\centering
\begin{tabular}{lcccccc}
\hline\hline\noalign{\smallskip}
SN & \multicolumn{6}{c}{Progenitor initial mass (\msun)} \\
   & \multicolumn{3}{c}{Non-rotating models} & \multicolumn{3}{c}{DBR models} \\
\cline{2-7}\noalign{\smallskip}
   & 16th & 50th & 84th & 16th & 50th & 84th \\ 
\hline\noalign{\smallskip}
2004A    & 12.1 & 13.4 & 15.0 & 11.5 & 12.8 & 14.5 \\
2004et   & 10.8 & 11.7 & 12.6 & 10.5 & 11.3 & 12.2 \\
2006my   & 12.0 & 14.6 & 18.0 & 11.5 & 13.7 & 16.8 \\
2008bk   &  9.0 &  9.1 &  9.7 &  9.0 &  9.0 &  9.7 \\
2008cn   & 15.4 & 16.7 & 18.5 & 14.5 & 16.0 & 17.5 \\
2009hd   & 18.2 & 19.7 & 21.3 & 16.5 & 18.6 & 21.2 \\
2009ib   & 14.6 & 17.1 & 20.1 & 13.8 & 16.3 & 19.4 \\
2009kr   & 13.5 & 17.4 & 22.3 & 12.6 & 16.5 & 21.5 \\
2009md   &  9.0 &  9.0 & 10.8 &  9.0 &  9.0 & 10.5 \\
2012A    &  9.0 &  9.5 & 10.4 &  9.0 &  9.0 & 10.5 \\
2012aw   & 12.1 & 13.7 & 15.6 & 11.5 & 13.2 & 15.1 \\
2012ec   & 16.5 & 18.2 & 19.6 & 15.4 & 17.2 & 18.7 \\
2013ej   &  9.9 & 10.7 & 11.5 &  9.7 & 10.5 & 11.4 \\
2017eaw  & 15.2 & 16.3 & 18.1 & 14.2 & 15.5 & 17.3 \\
2018zd   & 10.8 & 12.1 & 13.4 & 10.5 & 11.5 & 12.9 \\
2018aoq  &  9.8 & 11.0 & 12.6 &  9.4 & 10.6 & 12.2 \\
2022acko &  9.0 &  9.0 &  9.0 &  9.0 &  9.0 &  9.0 \\
2023ixf  & 13.2 & 14.2 & 15.3 & 12.5 & 13.5 & 14.6 \\
2024ggi  & 13.0 & 13.7 & 14.5 & 12.3 & 13.2 & 14.0 \\
2024abfl &  9.0 &  9.4 & 10.1 &  9.0 &  9.0 & 10.2 \\
2025pht  & 13.6 & 15.0 & 16.5 & 12.8 & 14.3 & 15.8 \\
\hline
\end{tabular}
\tablefoot{The results are given as the 16th, 50th, and 84th percentiles of the posterior distribution.}
\end{table}

\begin{figure}
\centering
\includegraphics[width=0.49\textwidth]{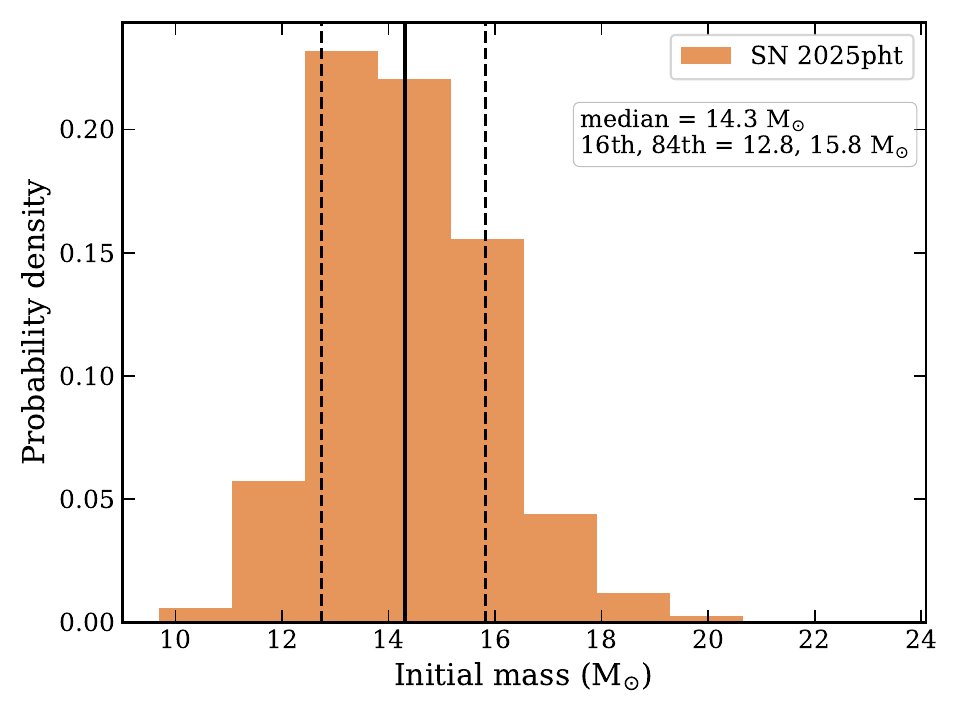}
\caption{Initial-mass distribution of the progenitor of SN~2025pht using the DBR models. The solid vertical line corresponds to the median of the distribution, while dashed vertical lines are the 16th and 84th percentiles.}
\label{fig:sn2025pht_mass_dist}
\end{figure}

We determined the probability distributions of the initial mass for each \snii\ progenitor in the sample using the MC method described in Sect.~\ref{subsec:mc}, combined with the stellar evolutions models presented in Sect.~\ref{subsec:models}.   
For comparison, we also computed results using only non-rotating models\footnote{For non-rotating models, the use of the \texttt{cKDTree} algorithm is unnecessary, as linear interpolation is sufficient.}.  
This allows us to directly assess the differences with respect to the DBR models, which constitute the main aim of this work.

As an example of our results, Fig.~\ref{fig:sn2025pht_mass_dist} shows the initial-mass distribution of the progenitor of SN~2025pht using the DBR models (i.e. including the initial surface velocity distribution and rotating stellar models). 
SN~2025pht is the most recent \snii\ with a detected progenitor in pre-explosion images, and the first to be identified with the \textit{James Webb} Space Telescope \citep{kilpatrick+25}.
We derived an initial mass of \mzams\eq14.3\ml1.5\,\msun, consistent with previous estimates \citep{kilpatrick+25}.

Table~\ref{table:results} lists the 16th, 50th, and 84th percentiles of the posterior distribution of the progenitor initial mass for each \snii\ in the sample. Results are shown both for the non-rotating and DBR models.
Additionally, the initial-mass distributions for the entire \snii\ sample are shown in Appendix~\ref{app:distributions} (Figs.~\ref{fig:mass_dist1} and \ref{fig:mass_dist2}).

We note that for the progenitor of SN~2022acko the resulting distributions, both from non-rotating and DBR models, are (almost) entirely concentrated at an initial mass value of 9\,\msun. This behaviour arises because the observed progenitor luminosity is significantly lower than the minimum luminosity covered by our models.
As described in Sect.~\ref{subsec:mc}, if during the MC analysis the progenitor luminosity falls below the minimum in the models, we set the progenitor initial mass to the lowest grid value, which is 9\,\msun. 
For the DBR models, all 10\,000 MC trials fall below this luminosity threshold. In the case of non-rotating models, only $\sim$1\% of the trials lie above the minimum model luminosity, resulting in a slightly broader distribution.
Although the progenitor luminosity lies below the range covered by our models, it could still be consistent with a CCSN at the lower initial mass limit, or with an ECSN \citep{vandyk+23b}.
We note that post-explosion images, once the SN emission has faded sufficiently, are required to confirm the progenitor through its disappearance.

A similar situation is found for SNe~2008bk, 2009md, 2012A, and 2024afbl, where part of the progenitor luminosity distribution extends below the minimum luminosity of the model grid. This results in the corresponding mass distributions being clustered around the lowest grid value.
In Sect.~\ref{subsec:low_luminosity_prog}, we discuss these low-luminosity progenitors in the context of stellar evolution models.

\section{Analysis}
\label{sec:analysis}

\begin{figure}
\centering
\includegraphics[width=0.49\textwidth]{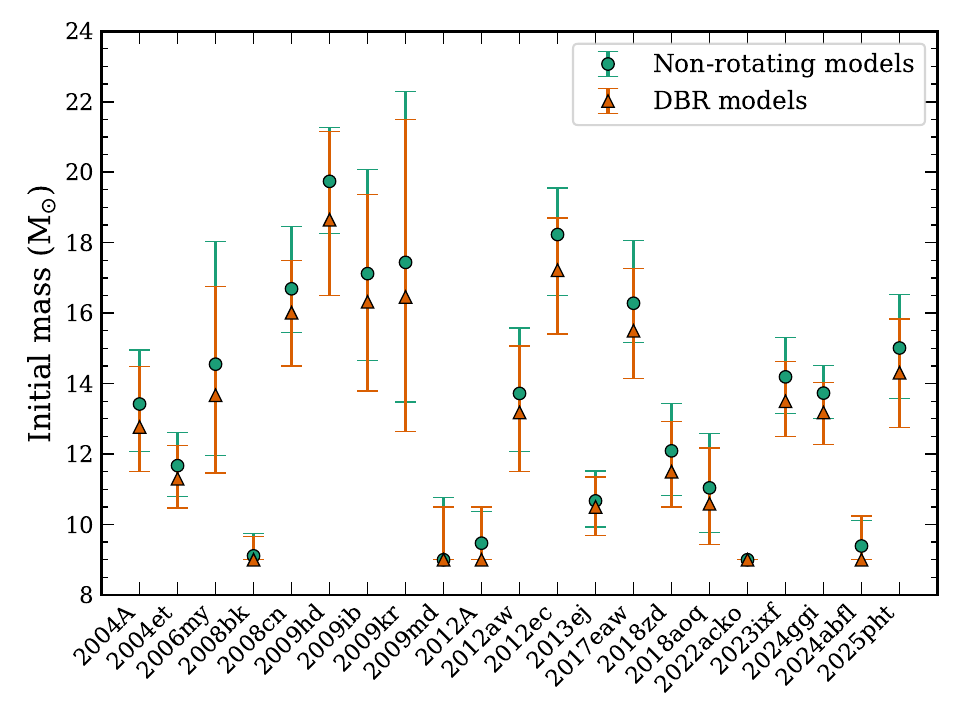}
\caption{Median of the initial-mass distribution of \snii\ progenitors in the sample, derived from non-rotating (green dots) and DBR models (orange triangles). Error bars correspond to the 16th and 84th percentiles.}
\label{fig:initial_mass_comp}
\end{figure}

\begin{figure}
\centering
\includegraphics[width=0.49\textwidth]{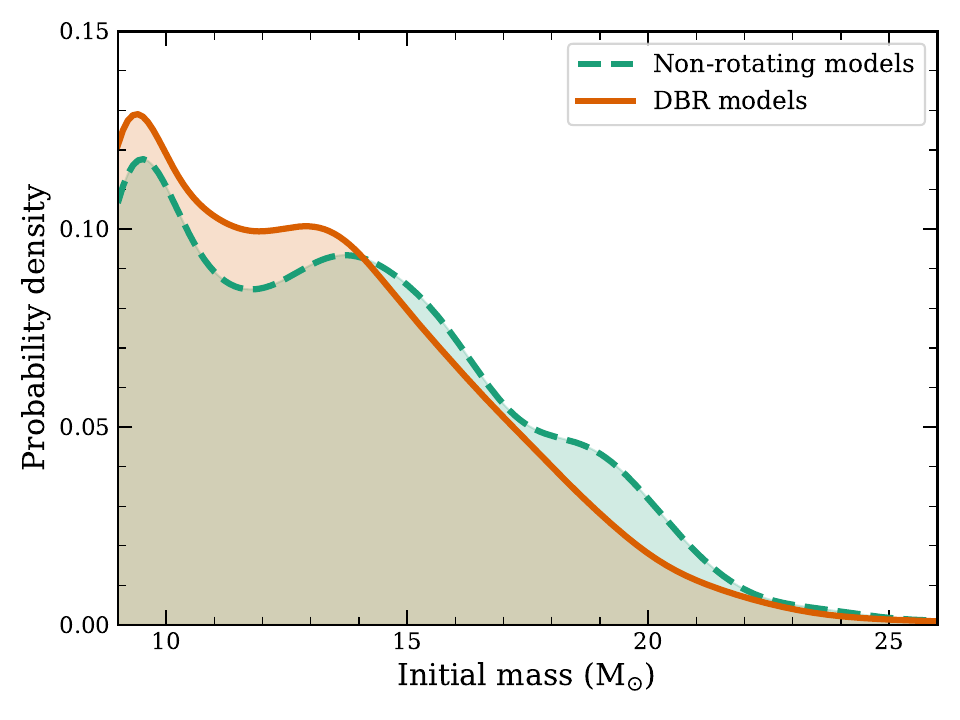}
\caption{Probability density functions of the \snii\ progenitor initial mass, derived from non-rotating (dashed green line) and DBR models (solid orange line).}
\label{fig:kde}
\end{figure}

\begin{figure}
\centering
\includegraphics[width=0.49\textwidth]{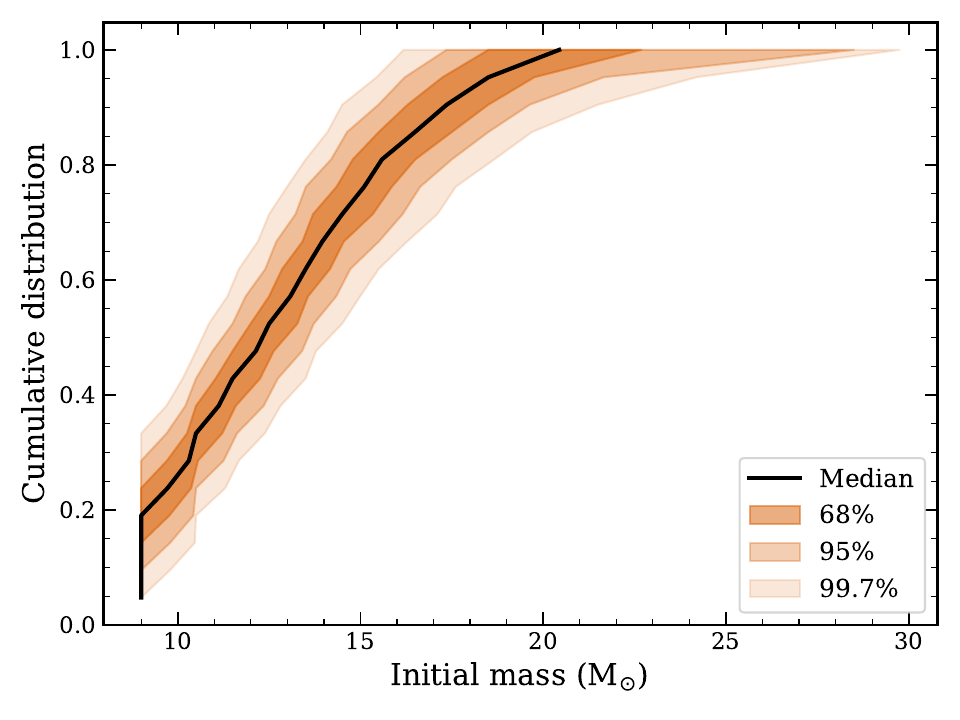}
\caption{Cumulative distributions of initial masses for the \snii\ progenitors in the sample derived from DBR models.}
\label{fig:cdf_rot}
\end{figure}

Having derived individual progenitor masses for each \snii\ in the sample from non-rotating and DBR models, we now perform a statistical analysis of the complete sample. This analysis includes comparisons between the two sets of estimates for each SN, the construction of probability density functions (PDFs) and cumulative distribution functions (CDFs), and the determination of the upper mass boundary.

Figure~\ref{fig:initial_mass_comp} compares the median values of the initial-mass distributions obtained with non-rotating and DBR models for each \snii\ progenitor in the sample, with error bars corresponding to the 16th and 84th percentiles. In all cases, the median masses inferred from DBR models are systematically lower than those from non-rotating models.
This systematic offset arises from one of the main effects of stellar rotation: enhanced chemical mixing. In rotating models, the increased mixing supplies additional material to the burning core, producing more massive cores. As a result, stars are more luminous throughout their evolution, including the pre-SN stage considered here. This luminosity increase explains the shift toward lower initial masses when adopting DBR models (see also Figs.~\ref{fig:mass_dist1} and \ref{fig:mass_dist2}). Although this trend was expected \citep[e.g.][]{straniero+19}, in the following we quantify the differences.

The smallest differences are found for SNe~2008bk, 2009md, 2012A, and 2024afbl. As discussed in Sect.~\ref{sec:results}, their inferred progenitor masses are clustered near the lower boundary of the model grid.
Excluding these cases, the smallest difference is 0.2\,\msun\ for SN~2013ej, while the largest is 1.1\,\msun\ for SN~2009hd, with the rest of the sample falling within this range.
Thus, the typical range of differences is $\sim$0.2$-$1.1\,\msun\ (Fig.~\ref{fig:initial_mass_comp}).
Moreover, although the distributions obtained with DBR models are shifted toward lower initial masses, the median values always remain within 1$\sigma$ of those derived from non-rotating models.
In general, the shifts in the inferred initial masses are modest, and the distributions largely overlap, indicating that the impact of including rotation is small compared to the typical observational uncertainties.

In the preceding analysis, we compared the initial-mass distributions for each \snii\ progenitor individually using both procedures.  
We now examine the PDF of the initial mass by considering all distributions simultaneously.  
To this end, we estimated the corresponding distribution using a Gaussian kernel density estimator applied to the MC samples described above, implemented with the \texttt{gaussian\_kde} function from the \texttt{SciPy} package.
Figure~\ref{fig:kde} shows the resulting PDFs for the non-rotating and DBR models. Although the distribution derived from the DBR models is shifted towards lower initial masses, the differences remain small.
To quantify the similarity between the two probability distributions, we computed the Hellinger distance \citep{hellinger09}. This metric takes values $H$ between 0 and 1, with $H\approx0$ indicating nearly identical distributions and $H\approx1$ corresponding to very different ones.
We obtain $H$\eq0.06, confirming that the two distributions are indeed very similar, consistent with the visual comparison from Fig.~\ref{fig:kde}.

We next focus on the lower and upper initial-mass boundaries of the distribution (\mlow, \mhigh). These are key properties commonly derived in \snii\ progenitor studies, which provide important constraints on the evolutionary pathways of massive stars \citep[e.g.][]{smartt15,davies+20a}.
To this end, we first constructed the CDF of the progenitor initial masses in our sample, following the approach of \citet{davies+20a}.
For each of the 10\,000~MC trials carried out in Sect.~\ref{sec:analysis}, we took the initial mass corresponding to each \snii\ progenitor in the sample and sorted the values in ascending order. This procedure allows us to construct the probability distribution of the CDF (see \citealt{martinez+22b}, for details).

Figure~\ref{fig:cdf_rot} presents the CDF derived from the DBR models, together with the corresponding confidence regions. For comparison, Fig.~\ref{fig:cdf} shows the results obtained with non-rotating models.
Our analysis suggests a minimum initial mass of \mlow\eq9\,\msun\ (although see the discussion in Sect.~\ref{subsec:low_luminosity_prog}).
Additionally, we obtain \mhigh\eq20.8$^{+2.8}_{-1.3}$\,\msun\ when using the non-rotating models, while for the DBR models we find \mhigh\eq20.4$^{+2.3}_{-1.9}$\,\msun. 
Thus, although the DBR models again yield a lower value than the non-rotating case, the difference is smaller than the uncertainty on \mhigh.

In summary, accounting for the observed distribution of stellar velocities leads to systematically lower progenitor initial mass estimates, but the effect remains modest compared to the statistical uncertainties. 
The \snii\ progenitor initial-mass distributions and the upper mass limit inferred from the sample are consistent with those obtained from non-rotating models.

\section{Discussion}
\label{sec:discussion}

\subsection{Low-luminosity progenitors}
\label{subsec:low_luminosity_prog}

Evolutionary models of \snii\ progenitors generally predict final luminosities above $\log L$~$\simeq$~4.5 (\citealt{eldridge+04, poelarends+08, woosley+15}, this work).
However, a number of \snii\ progenitors have been observed at lower luminosities ($\log L$~$\simeq$~4.3$-$4.5). Given that the observed progenitor luminosities lie below the minimum final luminosities reached by our models, the inferred mass distributions accumulate at the minimum initial mass of the grid, explaining the distributions inferred for SNe~2008bk, 2009md, 2012A, 2022acko, and 2024afbl (see Sect.~\ref{sec:results}).
This, in turn, indicates that a non-negligible fraction of \snii\ progenitors may populate the faint end of the luminosity distribution, which is not fully captured by current stellar evolution models.

The late evolutionary stages of stars close to the boundary between white-dwarf formation and SN explosions are particularly difficult to model. 
At solar metallicity, this transition occurs over a relatively narrow initial-mass range from 7 to 10\,\msun, where electron degeneracy and neutrino cooling play a dominant role in shaping the internal structure. These effects can lead to temperature inversions, off-centre shell ignition, and composition gradients, making the post-carbon-burning evolution particularly challenging to follow \citep[see e.g.][for details]{miyaji+80, nomoto+84, poelarends+08, woosley+15, limongi+24}.

In standard stellar evolution scenarios, stars in the initial-mass range of $\sim$7$-$10\,\msun\ form ONe cores following carbon burning and subsequently enter a thermally pulsating phase. These objects are known as super-asymptotic giant branch (SAGB) stars, and their evolutionary outcome depends on the balance between ONe core growth and mass loss, resulting in either an ONe white dwarf or an ECSN.
Stars with initial masses of $\sim$10$-$12\,\msun\ ignite neon off-centre, while more massive stars ignite neon centrally; in both cases, the evolution ultimately leads to a CCSN.
The exact mass limits remain uncertain, as they depend on the treatment of internal mixing, opacities, nuclear reaction rates, neutrino cooling, and numerical resolution, among other factors. In addition, these initial-mass limits  vary with initial metallicity and rotational velocity.

For stars that do not undergo the second dredge-up, there is an increasing relation between initial mass and final luminosity. However, when the second dredge-up occurs, the convective envelope penetrates deeply into the interior efficiently dredging up the products of previous nuclear burning stages.
This process precedes the onset of thermal pulses and results in a significant increase in the stellar luminosity, leading to values well above the minimum luminosity expected for CCSN progenitors \citep[e.g.][]{poelarends+08, fraser+11}.
Moreover, this leads to luminosities well above those inferred for some of the faintest observed \snii\ progenitors.
\citet{fraser+11} addressed the discrepancy between the low luminosities inferred for some \snii\ progenitors and predictions from stellar evolution models by proposing that stellar evolution proceeds sufficiently rapidly for the star to explode before the second dredge-up occurs.
To achieve this, they increased the cross section of the $^{12}$C~+~$^{12}$C reaction, motivated by the possible presence of a resonance \citep[see][for an updated discussion of this nuclear reaction]{monpribat+22}.
This approach allows progenitor models to reach luminosities as low as $\log L$~$\simeq$~4.3, thereby alleviating the tension between stellar evolution models and the observed luminosities of \snii\ progenitors.
Overall, significant uncertainties remain in the late evolutionary stages of low-mass CCSN progenitors, and the ability of stellar evolution models to reproduce the faint end of the \snii\ progenitor population remains an open question.

\subsection{The RSG problem}
\label{subsec:rsg_problem}

The first direct detection of a \snii\ progenitor was achieved more than two decades ago for SN~2003gd \citep{smartt+04}. Since then, the number of identified \snii\ progenitors has steadily increased, with more than 20 progenitors now detected, along with a comparable number of upper limits on progenitor luminosities \citep[see][for a recent review]{vandyk25}.
Early analyses of the \snii\ progenitor population based on direct progenitor detections suggested an upper initial-mass limit of $\sim$18\,\msun\ \citep{smartt+09, smartt15}.
This result appears to be in tension with stellar evolution models, which predict that more massive stars can retain substantial hydrogen-rich envelopes and thus produce \snii\ explosions, as well as with observations of more luminous ---and therefore more massive--- RSGs \citep{levesque+06}.
This absence of progenitors with initial masses above $\sim$25\,\msun\ is commonly referred to as the `RSG problem' \citep{smartt+09}.

\citet{davies+18} revised bolometric corrections and extinction estimates, finding an increased upper initial-mass limit of \mhigh\eq19\,\msun, with a 95\% upper confidence limit of <\,27\,\msun, corresponding to a statistical significance of the RSG problem at the $\sim$2$\sigma$ level; this result was later supported by \citet{davies+20a}, who analysed the progenitor population directly in luminosity space, thereby avoiding uncertainties associated with converting luminosities into initial masses using stellar evolution models.
Moreover, \citet{beasor+25} analysed the impact of systematic uncertainties and found that, even the most luminous RSGs remain consistent with the observed \snii\ progenitor sample at the $3\sigma$ level.
The reader is referred to \citet{morozova+18} and \citet{martinez+22b} for estimates of the statistical significance of the RSG problem based on light-curve modelling, and to \citet{fang+25} for estimates based on nebular spectral analyses.

Having determined the initial-mass distributions using both non-rotating and DBR models in Sect.~\ref{sec:analysis}, we now focus exclusively on the results from the DBR models.
These results indicate that the upper initial-mass limit for \snii\ progenitors is \mhigh\eq20.4$^{+2.3}_{-1.9}$\,\msun\ (68\% confidence), with a 95\% upper confidence limit of 28.5\,\msun. This corresponds to a statistical significance of the RSG problem at the $\sim$2$\sigma$ level.
We note, however, that the stellar evolution models used in this analysis were computed assuming solar metallicity. Adopting specific metallicity values to each individual SN could lead to variations in the inferred masses.
We emphasise that the analysis of the RSG problem constitutes an additional, exploratory analysis motivated by the availability of our results. The primary goal of this work is to compare the results obtained using non-rotating models and models computed with an observed distribution of initial rotational velocities.

\subsection{\snii\ progenitors from mass gainers reaching critical rotation}
\label{subsec:mass_gainers}

\begin{figure}
\centering
\includegraphics[width=0.49\textwidth]{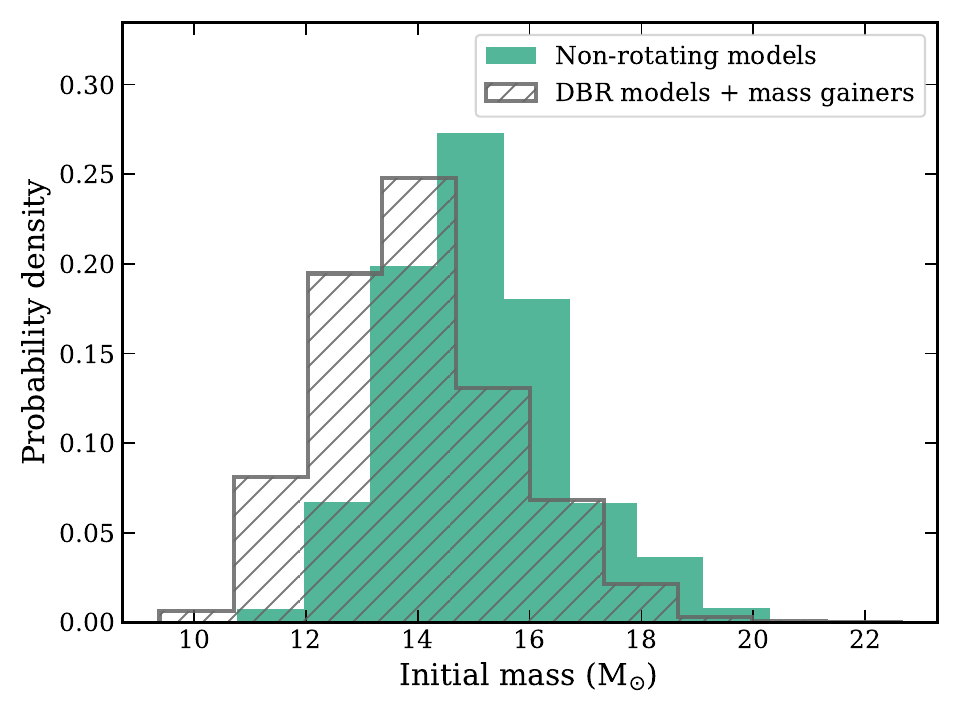}
\caption{Initial-mass distributions of the progenitor of SN~2025pht using non-rotating models (green solid bars) and DBR models including the contribution of mass gainers (grey dashed bars).}
\label{fig:sn2025pht_crit}
\end{figure}

In this work, we have focused exclusively on the single evolutionary channel for \snii\ progenitors.
However, it is well established that a large fraction of massive stars belong to binary systems (or higher-order multiples; \citealt{moe+17}) and are expected to interact with a companion during their evolution \citep{sana+12}.

In interacting binary systems, the initially more massive star transfers not only mass but also angular momentum to its companion.
The accretion of angular momentum can strongly affect the rotational properties and the evolution of the secondary star, allowing mass gainers to approach critical rotation after accreting relatively small amounts of material \citep{packet81, wang+20}.
Observationally, the distribution of rotational velocities of O- and B-type stars reveals a high-velocity tail, with projected velocities reaching up to $\sim$500\,km\,s$^{-1}$, commonly interpreted as the result of binary interaction \citep{de_mink+13}.

The accreted angular momentum spins up the outer layers of the secondary, and is subsequently transferred deeper into the star.
As a consequence, by the end of the binary interaction phase, the mass gainer is more massive than initially and rotates nearly uniformly approaching the critical surface velocity.
If this star subsequently undergoes advanced nuclear burning stages and retains a significant hydrogen-rich envelope at the end of its evolution, it may also explode as a \snii, thereby contributing to the observed \snii\ progenitor population.
In the following, we incorporate this evolutionary channel into our analysis to provide a first-order estimate of its impact on our results.

We computed additional evolutionary models of single massive stars at solar metallicity, initially rotating uniformly at their critical surface velocity. The initial-mass range spans 9$-$16\,\msun\ in steps of 1\,\msun. As discussed earlier, rotation enhances chemical mixing, producing more massive cores and, consequently, more luminous stars with stronger winds, which further reduce their hydrogen-rich envelopes. Therefore, models of more massive stars lose nearly all of their hydrogen-rich envelopes and therefore do not produce the SN type considered in this work.

To account for the mass-gainer evolutionary channel, we require an estimate of the fraction of \sneii\ expected to originate from it.
For this purpose, we adopted the results of \citet{zapartas+19}, who used binary population-synthesis simulations to estimate the fraction of \sneii\ arising from different evolutionary channels.
They find that 55\% of \snii\ progenitors originate from effectively single stars, i.e. stars that were born single or in very wide binary systems that do not experience binary interaction. In addition, about 14\% of \snii\ progenitors originate from stars that have accreted mass through stable Roche-lobe overflow from a companion. 
The remaining fraction arises from various merger channels, which are not considered in the present study.

To assess the impact of the mass-gainer evolutionary channel, we used the progenitor of SN~2025pht as a case study. We performed an analysis analogous to that described in Sect.~\ref{subsec:mc}.
First, we sampled progenitor luminosities from a Gaussian distribution (see Sect.~\ref{subsec:mc}).
For 55\% of the total MC trials, we assumed that the progenitors originate from the single-evolution channel and followed the same procedure as in the previous analysis using the DBR models.
For 14\% of the MC trials, we inferred the initial mass using the sampled luminosity and evolutionary models initially rotating at the critical surface velocity.
Merger scenarios were not considered.
We repeated this MC procedure 20\,000 times.

As a result, we obtained the initial-mass distribution for the progenitor of SN~2025pht including the contribution of the mass-gainer evolutionary channel, as shown in Fig.~\ref{fig:sn2025pht_crit}. 
We derived an initial mass of \mzams\eq13.7$^{+1.8}_{-1.3}$\,\msun. Although the inferred initial mass is lower than that obtained with the DBR models, a comparison with the non-rotating models shows that the median value remains consistent within 1$\sigma$.

This analysis provides a preliminary assessment of the impact of the mass-gainer evolutionary channel. We note, however, that the fraction of progenitors associated with each evolutionary channel is subject to uncertainties in stellar evolution modelling \citep[see][]{zapartas+19}.
We also assume that all mass gainers reach nearly uniform rotation at the critical surface velocity. 
Furthermore, in interacting binaries, mass transfer usually starts after the accreting star has already burned some fraction of its central hydrogen. In contrast, our critically rotating models start their evolution at the ZAMS. This temporal offset may affect the subsequent evolutionary phases.
Overall, while highly simplified, the inclusion of a mass-gainer channel highlights the impact of binary interaction on the properties of \snii\ progenitors and on the interpretation of progenitor mass distributions.

\subsection{Effect of metallicity on the progenitor initial-mass distribution}
\label{subsec:z_effect}

Metallicity is a fundamental parameter affecting stellar evolution and therefore the properties of SN progenitors. In the present study, however, we restrict our analysis to stellar models computed at solar metallicity.
An important effect of metallicity is its influence on stellar wind mass-loss rates. Given that the winds of hot stars are line-driven, with radiative acceleration primarily provided by UV metal lines (e.g. C, N, O and Fe-group elements), they are well known to depend on metallicity \citep[e.g.][]{vink+01, vink22}.
Such metallicity-driven differences in mass-loss behaviour directly affect the final properties of SN progenitors.
By contrast, on the luminous and cool side of the HR diagram, recent studies provide no evidence for an explicit dependence of RSG mass loss on metallicity, although this remains under study \citep{antoniadis+25, vanloon25}.

In massive stars, hydrogen burning proceeds predominantly via the CNO cycle. The cycle requires the presence of C, N, or O isotopes, which act as catalysts in the reaction network. 
Consequently, at lower metallicity, the initial abundance of these elements is reduced, which directly affects the energy generation of the CNO cycle. For a given density and temperature it will be lower than in the case of solar abundances.
This leads to changes in the central temperature and density during hydrogen burning, ultimately affecting the main-sequence lifetime, core masses, and therefore the final luminosities \citep{mowlavi+98}, which is the observable considered in this work.

The \texttt{GENEVA} and \texttt{MIST} grids of stellar evolution models at different initial metallicities show that, at fixed initial mass, the final stellar luminosity increases as the metallicity decreases \citep{ekstrom+12, georgy+13, choi+16, eggenberger+21}.
In other words, if the progenitor of a \snii\ has a subsolar initial metallicity, adopting solar-metallicity models to construct the initial-mass distribution will result in an overestimation of its mass, while the opposite occurs for supersolar metallicities.
This illustrates the impact of metallicity on the inferred progenitor masses. A more complete exploration of the parameter space, including variations in metallicity and rotation, will be considered in future work.

\section{Conclusions}
\label{sec:conclusions}

In this work, we quantified the differences in the inferred initial masses of \snii\ progenitors with pre-explosion detections when using non-rotating evolutionary models (a common approach in the literature) compared to rotating models, where the initial rotational velocities were drawn from the observed distribution of massive stars.
Using both approaches, we compared the progenitor initial masses for each SN, statistical distributions of the initial masses, and the upper initial mass limit.
In all cases, we find that the differences are small, and the values obtained with non-rotating models are consistent with those derived from DBR models.

Using the observed distribution of initial rotational velocities for \snii\ progenitors, we revisited the upper initial-mass limit inferred from pre-explosion imaging.
We obtained \mhigh\eq20.4$^{+2.3}_{-1.9}$\,\msun, with a 95\% upper confidence limit of 28.5\,\msun, which places the statistical significance of the RSG problem within the 2$\sigma$ level, in agreement with recent studies \citep{davies+20a, beasor+25}.

Our analyses indicate that, although rotation affects the evolutionary tracks of massive stars, its impact on progenitor mass estimates from pre-SN imaging remains modest within current uncertainties when the observed distribution of initial rotational velocities is taken into account. Therefore, we conclude that explicitly incorporating this distribution is not necessary to obtain significantly more precise estimates of progenitor initial masses.
These results offer a useful reference for future studies, suggesting that non-rotating models remain sufficient for estimating \snii\ progenitor masses, at least within current observational limits.

\section*{Data availability}

The tables containing the stellar properties at core carbon depletion for the full grid of models, along with the \texttt{MESA} input files used in this work, are publicly available at \url{https://doi.org/10.5281/zenodo.19099684}.

\begin{acknowledgements}
We thank the referee for a careful reading of the manuscript and for constructive comments that helped improve this work.
L.M. acknowledges support from a CONICET postdoctoral fellowship.
\textit{Software:} 
\texttt{MESA} \citep{paxton+11, paxton+13, paxton+15, paxton+18, paxton+19, jermyn+23},
\texttt{NumPy} \citep{numpyguide2006, numpy2011}, \texttt{matplotlib} \citep{matplotlib}, 
\texttt{SciPy} \citep{scipy},  
\texttt{pandas} \citep{pandas}, 
\texttt{jupyter} \citep{jupyter}.
\end{acknowledgements}

\bibliographystyle{aa.bst}
\bibliography{biblio.bib}

\begin{appendix}

\onecolumn
\begin{figure*}
\section{Hertzsprung--Russell diagrams}
\label{app:hr}

Figures~\ref{fig:hr_all1} and \ref{fig:hr_all2} show HR diagrams for a subsample of stellar models. Each panel corresponds to a fixed initial mass and different initial rotational velocities.

\centering
\includegraphics[width=0.92\textwidth]{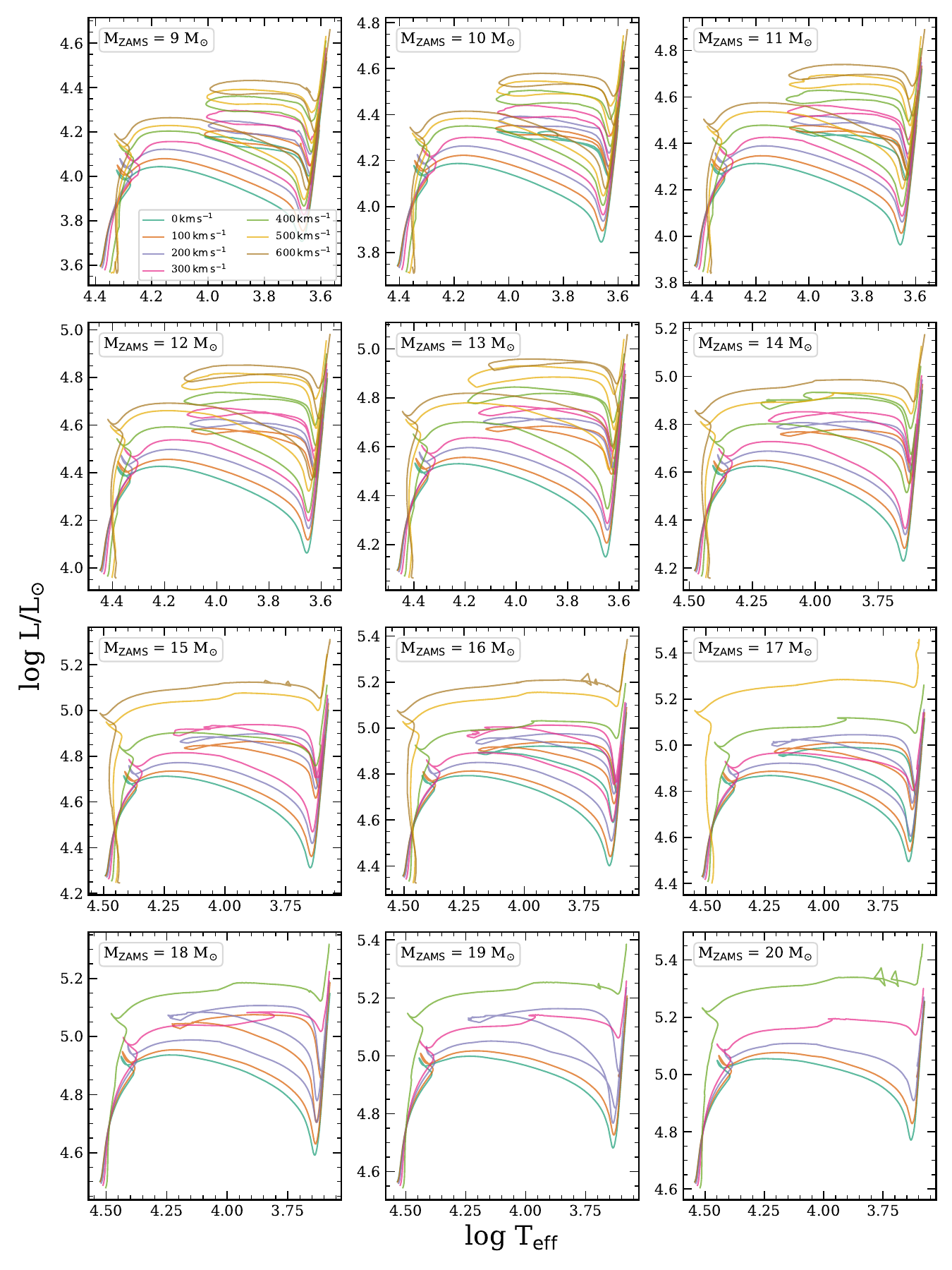}
\caption{HR diagrams for stellar evolutionary models. Each panel corresponds to a fixed initial mass, as indicated in the panel labels. For a given mass, evolutionary tracks computed with different initial rotational velocities are shown. Only a subset of models is shown for visualisation purposes.}
\label{fig:hr_all1}
\end{figure*}

\begin{figure*}
\includegraphics[width=0.92\textwidth]{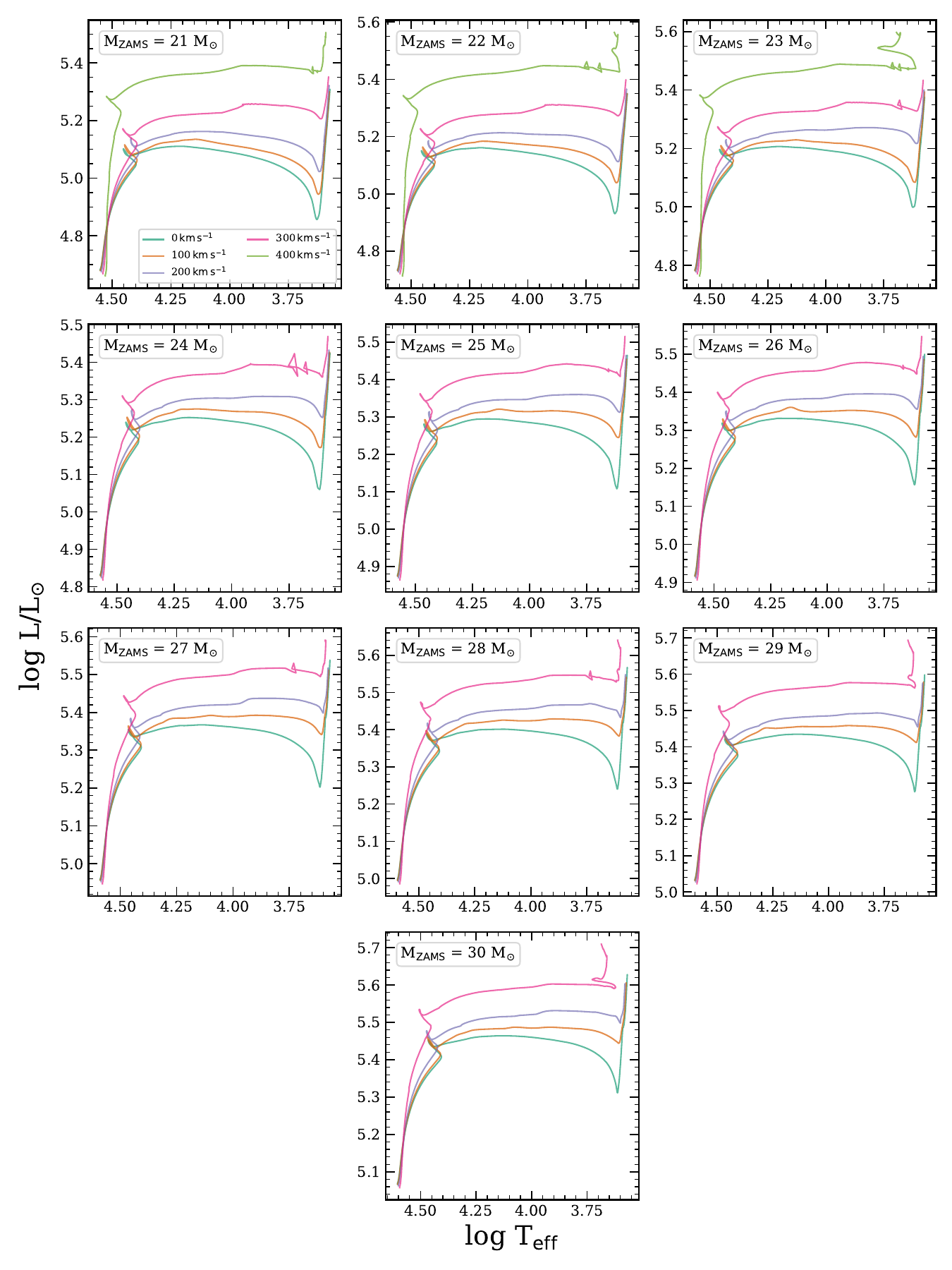}
\caption{continued.}
\label{fig:hr_all2}
\end{figure*}

\begin{figure*}

\section{Additional progenitor mass distribution diagrams}
\label{app:distributions}

Figures~\ref{fig:mass_dist1} and \ref{fig:mass_dist2} show the initial-mass distributions for the complete sample of \snii\ progenitors, obtained from non-rotating and DBR models.
Additionally, Fig.~\ref{fig:cdf} shows the cumulative distribution function of the initial mass for the \sneii\ in the sample, derived from non-rotating models.

\includegraphics[width=0.98\textwidth]{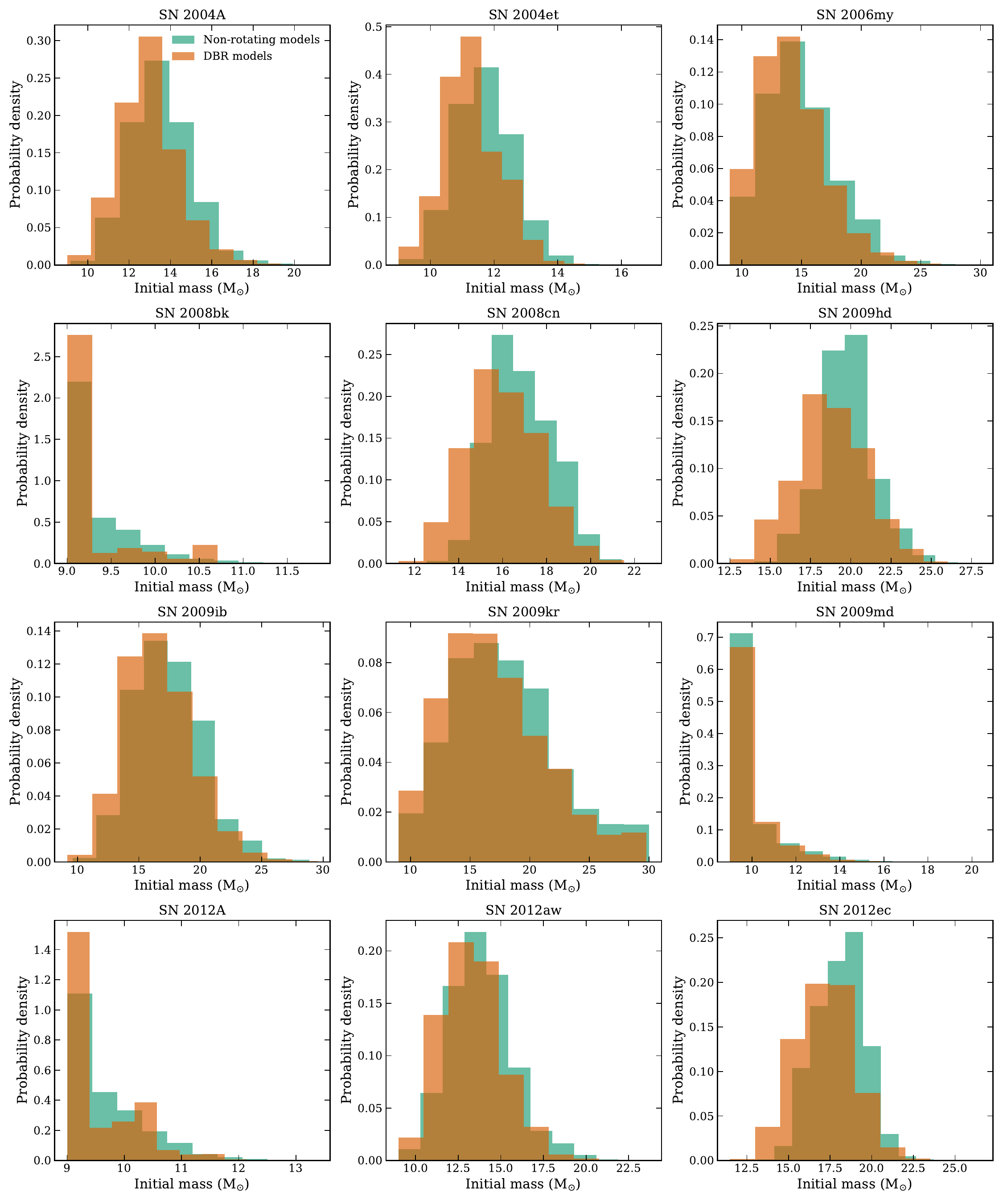}
\caption{Initial-mass distributions of \snii\ progenitors in the sample from non-rotating (green) and DBR models (orange).}
\label{fig:mass_dist1}
\end{figure*}

\begin{figure*}
\centering
\includegraphics[width=0.98\textwidth]{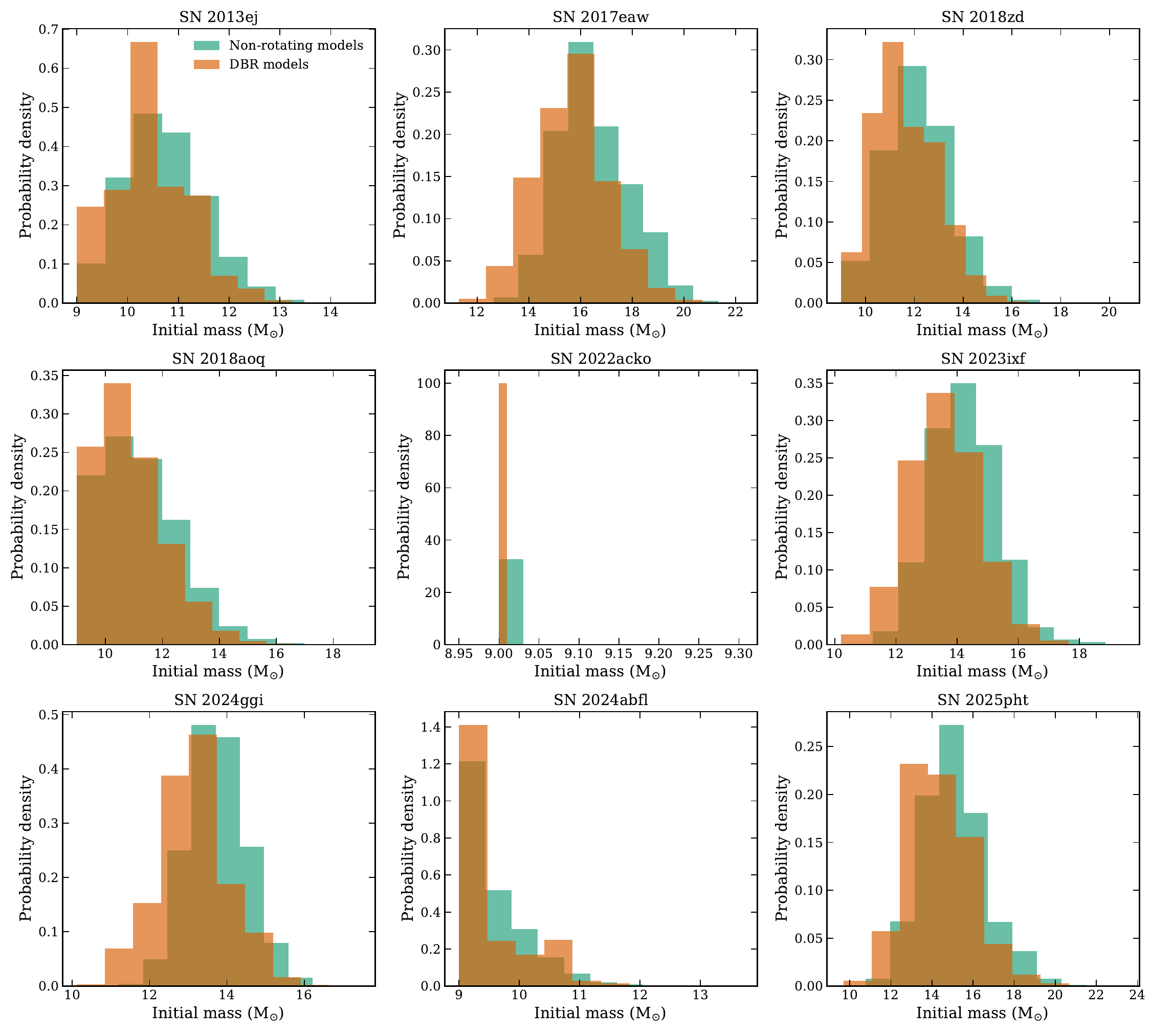}
\caption{continued.}
\label{fig:mass_dist2}
\end{figure*}

\begin{figure*}
\centering
\includegraphics[width=0.49\textwidth]{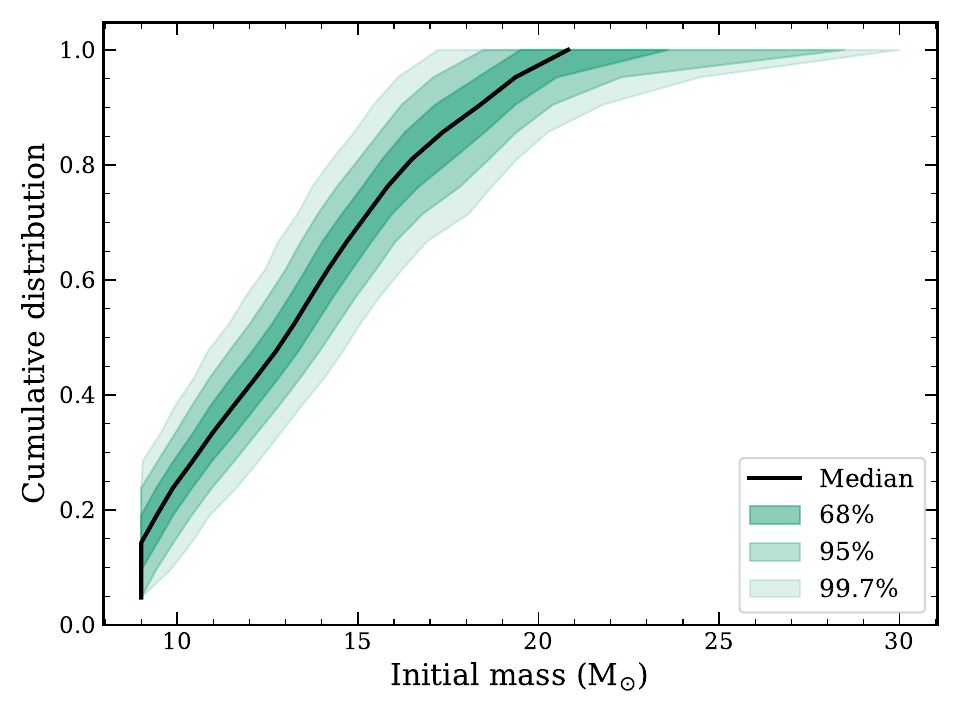}
\caption{Cumulative distributions of initial masses for the \snii\ progenitors in the sample derived from non-rotating models.}
\label{fig:cdf}
\end{figure*}

\end{appendix}

\end{document}